\documentclass{article}

\usepackage{arxiv}

\usepackage{xcolor}
\usepackage{amsthm}
\usepackage{tensor}
\usepackage{empheq}
\usepackage{amssymb}
\usepackage{enumitem}
\usepackage[utf8]{inputenc} 
\usepackage[T1]{fontenc}    
\usepackage{hyperref}       
\usepackage{url}            
\usepackage{booktabs}       
\usepackage{amsfonts}       
\usepackage{nicefrac}       
\usepackage{microtype}      
\usepackage{lipsum}
\usepackage{graphicx}
\graphicspath{ {./images/} }

\usepackage{tabularx}
\usepackage{array}
\usepackage{amsmath}

\title{Preserving the energy-momentum tensor in \texorpdfstring{$f(R,\text{Matter})$}{fRmatter} theories}

\author{
Szőllősi Tamás-Géza\thanks{
\texttt{szollosigeza@student.elte.hu}}\\
Faculty of Science, Eötvös Loránd University\\
Faculty of Informatics, Eötvös Loránd University\\
Budapest, Hungary
}

\begin{document}

\maketitle

\begin{abstract}
In certain modified theories of gravity, non-minimal couplings between matter and geometry lead to the nonconservation of the energy-momentum tensor. By interpreting this as an effective dissipative process, we formulate a general class of \(f(R,\text{Matter})\) theories with the Herglotz variational principle, a variational approach designed for dissipative systems. We demonstrate that, for an appropriate choice of the Herglotz contribution, the resulting Herglotz extension of \(f(R,\text{Matter})\) gravity restores the covariant conservation of the energy-momentum tensor.
\end{abstract}

\tableofcontents

\newpage

\section{Introduction}

Einstein did not arrive at the field equations of general relativity in a single step, but through a long process of trial and error guided by physical principles. A remarkable feature of the final theory is the covariant conservation of the energy-momentum tensor, which is a direct consequence of the contracted Bianchi identity. The Einstein field equations can also be derived from a variational principle based on the Einstein-Hilbert action supplemented by a matter action. Under the assumption of minimal coupling between matter and geometry, the diffeomorphism invariance of the matter Lagrangian implies the local conservation of the energy-momentum tensor \cite{landau1975}. It is important to note that the assumption of minimal coupling is crucial. Once this assumption is relaxed, the relation between diffeomorphism invariance and energy-momentum conservation becomes subtle, see for example \cite{universe11120386, BarroseSa:2025uxe}.

Non-conservative theories of gravity became appealing because they may provide a framework for describing dark energy phenomenology \cite{Sahlu2024CosmologyEPJC}, while also alleviating current cosmological tensions \cite{DUBEY2025116938}. A wide class of such theories arises from non-minimal couplings between matter and geometry, including, for example, $f(R,T), f(R,L_m)$ and $f(R,T_{\mu\nu} T^{\mu\nu})$ (for a review on non-conservative theories, see \cite{Akarsu2020RastallLCDM}). It is worth noting, however, that in some cases the non-conservative effects are almost negligible, and current observations show no statistically significant evidence for departures from general relativity \cite{Akarsu2020RastallLCDM}.

A second mechanism to generate non-conservative effects, besides introducing non-minimal couplings, is to break diffeomorphism invariance. This can be achieved, for example, by introducing a non-dynamical background field, as in Paiva's nonconservative gravity \cite{herglotz_mathematical,herglotz_examples,herglotz_GR1,herglotz_GR2}. Therein, the background field is mathematically interpreted as a closed one-form, and the field equations are derived from a generalization of the usual Hamiltonian variational principle, in which the Lagrangian is allowed to depend explicitly on the action itself. This generalized variational principle is known as the  Herglotz variational principle. 

Originally, the Herglotz principle was developed to provide a variational description for dissipative systems, but recently, it has proven to be useful in modified gravity theories. In \cite{HerglotzFRT} it was shown  that generalizing Paiva's action to $f(R,T)$ gravity gives a new extension of $f(R,T)$ gravity, in which the main criticisms regarding the impossibility of obtaining cosmologically viable solutions for models of the type $f(R,T)=R+\alpha T$ (due to the constant deceleration parameter) can be cured. In the scalar-tensor representation of the theory, a conservative cosmological model has also been found to be able to explain current observational data \cite{scalar_marek}.

The purpose of this paper is twofold: first, we introduce a systematic way of treating $f(R,T),f(R,\mathcal{L}_m),f(R,T_{\mu \nu} T^{\mu \nu})$ and other theories with non-minimal matter geometry coupling in a unified way, which we dub $f(R,\text{Matter})$ gravity. Then, we formulate this class of theories in the Herglotz variational framework, and show that the dynamics of the Herglotz field can be naturally chosen so as to preserve conservation of the energy-momentum tensor.

The paper is structured as follows. In section \ref{sec2}, we formulate $f(R,\text{Matter})$ gravity, derive its field equations, and show that generically the energy-momentum tensor is not covariantly conserved. As a consequence, test particles do not follow geodesics, and we determine the extra force explicitly. In section \ref{sec3}, we briefly review the Herglotz variational principle in classical mechanics and classical field theory, obtaining the field equations from a new perspective. In section \ref{sec4}, we formulate $f(R,\text{Matter})$ theories within the Herglotz variational framework and explicitly show that they generalize previous results in the literature. For completeness, we also present their scalar--tensor equivalents. In subsection \ref{sec5}, we show that energy-momentum conservation can be naturally achieved in all these theories, thereby providing several new classes of conservative theories within $f(R,T),f(R,L_m), f(R,T_{\mu\nu} T^{\mu\nu})$ gravity. Finally, we conclude the paper with a discussion and outlook for possible future research in section \ref{sec6}.

\section{\texorpdfstring{$f(R,\text{Matter})$}{fRmatter} gravity}
\label{sec2}
Consider a system with variables $\{g_{\mu\nu}, \Psi\}$, where $g_{\mu\nu}$ metric describes geometry and $\Psi$ collectively describes the matter fields, $\Psi:=\{\Psi_1,\dots,\Psi_{n} \}$, for each matter type. Working in natural units $c=G=1$, we consider an action of the form
\begin{equation}
    S[g_{\mu\nu},\Psi]=\int_{\mathcal{V}} \left[\frac{1}{16\pi}f(R,\Phi(g_{\mu\nu},\Psi,\partial_{\mu}\Psi))+\mathcal{L}_{m}(g_{\mu\nu},\Psi,\partial_{\mu}\Psi)\right]\sqrt{-g}d^{4}x,
    \label{eq1}
\end{equation}
where $f(R,\Phi)$ is an arbitrary analytic function of the Ricci scalar, $R$, and $\Phi$, a matter scalar invariant. $\mathcal{L}_{m}$ is the Lagrangian density describing matter, and we define the energy-momentum tensor as
\begin{equation}
    T_{\mu\nu}=-\frac{2}{\sqrt{-g}}\frac{\delta(\sqrt{-g}\mathcal{L}_{m})}{\delta g^{\mu\nu}}.
    \label{eq2}
\end{equation}
The field equations are obtained through extremizing the action
\begin{equation}
    \delta S[g,\Psi]=\delta \left( \int_{\mathcal{V}} \left[\frac{1}{16\pi}f(R,\Phi(g_{\mu\nu},\Psi,\partial_{\mu} \Psi))+\mathcal{L}_{m}(g_{\mu\nu},\Psi,\partial_{\mu}\Psi)\right]\sqrt{-g}d^{4}x\right)=0 .
\end{equation}

Varying with respect to the matter variables, we obtain from the Euler-Lagrange equations
\begin{equation}
    \frac{\partial \left(\frac{1}{16\pi}f(R,\Phi)+\mathcal L_{m}\right)}{\partial \Psi}
-\,\nabla_\mu\!\left(\frac{\partial \left(\frac{1}{16\pi}f(R,\Phi)+\mathcal L_{m}\right)}{\partial (\partial_\mu\Psi)}\right)=0.
\label{eq5}
\end{equation}
For minimally coupled theories, where $f$ does not depend on $\Phi$, Eq. \eqref{eq5} simplifies to
\begin{equation}
    \frac{\partial \mathcal L_{m}}{\partial \Psi}
-\,\nabla_\mu\!\left(\frac{\partial \mathcal L_{m}}{\partial (\partial_\mu\Psi)}\right)=0.
\label{eq6}
\end{equation}
The variation with respect to the metric gives
\begin{equation}
    \frac{\delta S}{\delta g^{\mu\nu}}=\int_{\mathcal{V}} [f_{R}(R,\Phi)R_{\mu\nu}+(g_{\mu\nu}\Box-\nabla_{\mu}\nabla_{\nu})f_{R}(R,\Phi)+f_{\Phi}(R,\Phi)\Xi_{\mu\nu}-\frac{1}{2}g_{\mu\nu}f(R,\Phi)-8\pi T_{\mu\nu}]\sqrt{-g}d^{4}x=0,
\end{equation}
where we introduced the notations
\begin{equation}
    f_{\Phi}:=\partial f/ \partial \Phi, \;f_{R}:=\partial f/ \partial R\quad \text{and} \quad\Xi_{\mu\nu}:=\frac{\delta \Phi}{\delta g^{\mu\nu}}
\end{equation}
for the variation of the matter scalar invariant with respect to the metric. Hence, the field equations of $f(R,\text{matter})$ gravity are
\begin{equation}
    f_{R}(R,\Phi)R_{\mu\nu}-\frac{1}{2}g_{\mu\nu}f(R,\Phi)+(g_{\mu\nu}\Box-\nabla_{\mu}\nabla_{\nu})f_{R}(R,\Phi)=8\pi T_{\mu\nu}-f_{\Phi}(R,\Phi)\Xi_{\mu\nu}.
    \label{eq8}
\end{equation}
It can be immediately seen that choosing $f_{\Phi}=0$, the field equations of $f(R)$ gravity are recovered \cite{Sotiriou:2008rp}.

Taking the divergence of Eq.\eqref{eq8}, with the help of the identity $(\Box\nabla_{\nu}-\nabla_{\nu}\Box)F=R_{\mu\nu}\nabla^{\mu}F$ \cite{koivisto}, we obtain for the divergence of the energy-momentum tensor the expression
\begin{equation}
    \nabla^{\mu}T_{\mu\nu}=\frac{1}{8\pi}\left[-\frac{1}{2}g_{\mu\nu}f_{\Phi}(R,\Phi)\nabla^{\mu}\Phi+\nabla^{\mu}(f_{\Phi}(R,\Phi)\Xi_{\mu\nu})\right],
    \label{eq9}
\end{equation}
from which we see that generally the energy-momentum tensor is not covariantly conserved. 

As a consequence, in $f(R,\text{Matter})$ gravity test particles moving in a gravitational field, do not follow geodesics \cite{harko_fR_T}. To prove this, first we assume that matter can be described as an ideal fluid, given by the energy-momentum tensor
\begin{equation}
    T_{\mu\nu}=(p+\rho)u_{\mu}u_{\nu}-g_{\mu\nu}p,
\end{equation}
where $u_{\mu}$ is the four-velocity, which  satisfies the condition $u_{\mu}u^{\mu}=1$.

Thus, Eq.\eqref{eq9} can be written in the form
\begin{equation}
    [\nabla_{\nu}(\rho+p)]u^{\mu}u^{\nu}+
(\rho+p)u^{\mu}\nabla_{\nu}u^{\nu}+(\rho+p)u^{\nu}\nabla_{\nu}u^{\mu}-g^{\mu\nu}\nabla_{\nu}p=\frac{1}{8\pi}\left[-\frac{1}{2}g^{\mu\nu}f_{\Phi}(R,\Phi)\nabla_{\nu}\Phi+\nabla_{\nu}(f_{\Phi}\Xi^{\mu\nu})\right].
\end{equation}

Following the same procedure as in \cite{harko_fR_T}, we introduce the projection operator $h_{\mu\lambda}:=g_{\mu\lambda}-u_{\mu}u_{\lambda}$, which obeys $h_{\mu\lambda}u^{\mu}=0$. Multiplying the expression above with the projection operator leads to
\begin{equation}(\rho+p)g_{\mu\lambda}u^{\nu}\nabla_{\nu}u^{\mu}-h_{\mu\lambda}g^{\mu\nu}\nabla_{\nu}p=\frac{h_{\mu\lambda}}{8\pi}\left[-\frac{1}{2}g^{\mu\nu}f_{\Phi}(R,\Phi)\nabla_{\nu}\Phi+\nabla_{\nu}(f_{\Phi}\Xi^{\mu\nu})\right],
\end{equation}
which upon contraction with the inverse metric $g^{\lambda\alpha}$ can be written as
\begin{equation}
    u^{\nu}\nabla_{\nu}u^{\alpha}=\frac{h_{\mu}{}^{\alpha}}{8\pi(\rho+p)}\left[\left(8\pi\nabla_{\nu}p-\frac{1}{2}f_{\Phi}(R,\Phi)\nabla_{\nu}\Phi\right)g^{\mu\nu}+\nabla_{\nu}(f_{\Phi}\Xi^{\mu\nu})\right].
\end{equation}
Taking into account the relation
\begin{equation}
    u^{\nu}\nabla_{\nu}u^{\mu}=\frac{d^2x^{\mu}}{ds^2}+\Gamma^{\mu}{}_{\nu\lambda}u^{\nu}u^{\lambda},
\end{equation}
we obtain the equation of motion of a test fluid in $f(R,\text{Matter})$ gravity 
\begin{equation}
    \frac{d^2x^{\mu}}{ds^2}+\Gamma^{\mu}{}_{\nu\lambda}u^{\nu}u^{\lambda}=\mathcal{F}^{\mu},
    \label{eq:EOM}
\end{equation}
where the extra-force takes the form
\begin{equation}\label{eq:force}
    \mathcal{F}^{\mu}:=\frac{h_{\alpha}{}^{\mu}}{8\pi(\rho+p)}\left[\left(8\pi\nabla_{\nu}p-\frac{1}{2}f_{\Phi}(R,\Phi)\nabla_{\nu}\Phi\right)g^{\alpha\nu}+\nabla_{\nu}(f_{\Phi}(R,\Phi)\Xi^{\alpha\nu})\right].
\end{equation}
Note that by setting $f_{\Phi}=0$,  we obtain the equation of
motion of perfect fluids in classical general relativity, i.e., 
\begin{equation}
    \frac{d^2x^{\mu}}{ds^2}+\Gamma^{\mu}{}_{\nu\lambda}u^{\nu}u^{\lambda}=\frac{h^{\mu\nu}\nabla_{\nu}p}{(\rho+p)}.
\end{equation}
Furthermore, in the limit $p\to 0$, which corresponds to a pressureless fluid (dust), the motion of the test particles becomes geodesic.

In the following sections, we will extend $f(R,\text{Matter})$ gravity using the Herglotz variational principle, and show that by imposing an appropriate condition on the Herglotz field, the extra-force in Eq. \eqref{eq:force} can be cancelled even for non-trivial couplings $f_{\Phi} \neq 0$.

\section{Herglotz variational principle}
\label{sec3}
In this section, we briefly review the Herglotz variational principle in classical mechanics, and classical field theory. In the field theory case, we use a similar formalism as in the mechanics case, contrary to previous works on the topic. We illustrate key concepts through representative examples, then pass to examples in gravity. 

\subsection{Herglotz variational principle in classical mechanics}

To naturally incorporate dissipation, Hamilton's principle is generalized as follows: instead of extremizing a fixed action functional, the Herglotz approach introduces an action functional whose evolution depends explicitly on the Lagrangian itself, through a differential equation.

Let $q(t)$ denote a generalized coordinate defined over the time interval $[t_{0},t_{1}]$ and $S(t)$ be the action functional. 
The Herglotz variational principle is defined by the differential equation
\begin{equation}
\dot S(t)=L\bigl(q(t),\dot q(t),S(t),t\bigr),
\label{eq18}
\end{equation}
with fixed initial condition $S(t_0)=S_0$ and fixed endpoints $q(t_0)$ and $q(t_1)$.

Taking the variation of Eq.\eqref{eq18}, we arrive at
\begin{equation}
    \delta \dot{S}=\frac{\partial L}{\partial q}\delta q+\frac{\partial L}{\partial \dot q}\delta \dot q+\frac{\partial L}{\partial S}\delta S.
    \label{eq19}
\end{equation}
In order to simplify our calculations, we introduce the parameter $\rho(t):=\exp\left(-\int_{t_0}^{t} \frac{\partial L}{\partial S}dt\right)$, with the help of which we can rewrite Eq.\eqref{eq19} as
\begin{equation}
    \left[\rho(t)\frac{\partial L}{\partial q}-\frac{d}{dt}\left(\rho(t)\frac{\partial L}{\partial \dot q}\right)\right]\delta q=0.
\end{equation}
Requiring stationarity of $S(t_1)$ under variations $\delta q(t)$ yields the generalized Euler-Lagrange equations
\begin{equation}
\frac{d}{dt}\left(\frac{\partial L}{\partial \dot q}\right)
-\frac{\partial L}{\partial q}
+\frac{\partial L}{\partial S}\,\frac{\partial L}{\partial \dot q}
=0.
\label{eq21}
\end{equation}
In the limit where the Lagrangian does not depend on $S$, i.e.\ $\partial L/\partial S=0$, 
Eq.\eqref{eq21} reduces to the standard Euler-Lagrange equations of classical mechanics.

A simple and illustrative example is obtained by considering a Lagrangian of the form
\begin{equation}
L(q,\dot q,S)=\frac{1}{2}m\dot q^{\,2}-V(q)+\lambda S,
\label{eq22}
\end{equation}
where $\lambda\in R^{*}$ is a constant. 
Substituting \eqref{eq22} into the generalized Euler-Lagrange equations \eqref{eq21} leads to
\begin{equation}
m\ddot q+\lambda m\dot q+\frac{dV}{dq}=0,
\end{equation}
which describes a particle subject to linear damping.


\subsection{Herglotz variational principle in classical field theories}

Let $(\mathcal{M},\eta)$ be the $n$-dimensional Minkowski spacetime. Consider a region $\mathcal{V}$ of the manifold with smooth boundary $\Omega$ and unit normal vector $n_{\mu}$. The Herglotz variational principle for fields can be formulated as
\begin{align}
&S=\int_{\Omega} s^\mu n_\mu \, d^{n-1}x=\int_{\mathcal{V}}\partial_{\mu}s^{\mu}d^{n}x \\
&\partial_\mu s^\mu=\mathcal L\bigl(x^\mu,\phi,\partial_\mu \phi,s^\mu\bigr),
\label{eq25}
\end{align}
with fixed boundary conditions $\phi(\Omega)=\phi^{\Omega}$, $\partial_\mu \phi(\Omega)=(\partial_\mu \phi)^{\Omega}$.The vector field $s^{\mu}$ plays the role of the action density, extending the notion of the classical action functional $S$ to the covariant framework.

Taking the variation of Eq.\eqref{eq25}, we obtain
\begin{equation}
\partial_\mu (\delta s^\mu)
=
\frac{\partial \mathcal L}{\partial \phi}\,\delta\phi
+
\frac{\partial \mathcal L}{\partial (\partial_\mu\phi)}\,
\delta(\partial_\mu\phi)
+
\frac{\partial \mathcal L}{\partial s^\mu}\,\delta s^\mu,
\label{eq24}
\end{equation}
which, by introducing the parameter $\rho(x^{\mu})$ defined by $\partial_{\mu}\rho:=-\rho\frac{\partial\mathcal{L}}{\partial s^{\mu}}$, can be rewritten as
\begin{equation}
\partial_\mu(\rho\,\delta s^\mu)
=
\rho
\left(
\frac{\partial \mathcal L}{\partial \phi}\,\delta\phi
+
\frac{\partial \mathcal L}{\partial (\partial_\mu\phi)}\,
\delta(\partial_\mu\phi)
\right).
\label{eq27}
\end{equation}
Integrating \eqref{eq27} over $\mathcal{V}$ yields
\begin{equation}
    \int_{\mathcal{V}}
\left[
\rho\,\frac{\partial \mathcal L}{\partial \phi}
-
\partial_\mu\!\left(
\rho\,\frac{\partial \mathcal L}{\partial (\partial_\mu\phi)}
\right)
\right]
\delta\phi
\, d^n x =0.
\end{equation}
After substitution we obtain the generalized Euler-Lagrange equations for fields
\begin{equation}
    \frac{\partial \mathcal L}{\partial \phi}
-\,\partial_\mu\!\left(\frac{\partial \mathcal L}{\partial (\partial_\mu\phi)}\right)+\frac{\partial \mathcal{L}}{\partial s^{\mu}}\frac{\partial \mathcal L}{\partial (\partial_\mu\phi)}=0.
\label{eq29}
\end{equation}

As an illustration of the formalism, consider the non-conservative formulation of the Klein-Gordon equation which can be obtained through the Lagrangian density
\begin{equation}
    \mathcal{L}=\frac{1}{2}\partial^{\mu}\phi\partial_{\mu}\phi-\frac{1}{2}m^{2}\phi^{2}+\lambda_{\mu}s^{\mu},
    \label{eq28}
\end{equation}
where $\lambda_{\mu}=(\lambda_{0},\lambda_{1},\lambda_{2},\lambda_{3})$ is a constant four-vector.

Using the generalized Euler-Lagrange equation \eqref{eq29} leads to
\begin{equation}
    (\Box+m^{2})\phi-\lambda_{\mu}\partial^{\mu}\phi=0.
    \label{eq31}
\end{equation}
Note that by considering the special case when $\lambda_{\mu}=(\lambda_{0},0,0,0)$, Eq.\eqref{eq31} reduces to the Telegraph equation.

\subsection{Herglotz variational principle in gravitational theories }
In this subsection, we present the application of the Herglotz variational principle to gravitational theories. First, we introduce action-dependent gravity, following Paiva \cite{herglotz_GR1,herglotz_GR2}, then in the next section we generalize the formalism to Herglotz-type $f(R,\text{Matter})$ theories, recovering as a special case the recently proposed Herglotz-type $f(R,T)$ theory \cite{HerglotzFRT}.

Let $(\mathcal{M},g)$ be an $n$-dimensional smooth manifold with a Lorentzian metric $g$. Consider a region $\mathcal{V}$ of the manifold with smooth boundary $\Omega$, unit normal vector $\eta_{\mu}$ and induced metric $h$. The Herglotz variational principle for gravitational theories is given by
\begin{equation}
S(\Omega)=\int_{\Omega}\eta_{\mu}s^{\mu}\sqrt{-h}d^{n-1}x=\int_{\mathcal{V}}\nabla_{\mu}s^{\mu}\sqrt{-g}d^{n}x.
    \label{eq32}
\end{equation}
\begin{equation}
        \nabla_{\mu}s^{\mu}=\mathcal{L}(x^{\mu},g_{\mu\nu},\partial_{\lambda}g_{\mu\nu},s^{\mu}),
        \label{eq33}
\end{equation}
with fixed boundary conditions $g_{\alpha\beta}(\Omega)=g_{\alpha\beta}^\Omega$ and $\partial_{\sigma}g_{\alpha\beta}(\Omega)=(\partial_{\sigma}g_{\alpha\beta})^{\Omega}$.

To illustrate the formalism, we present Paiva's action-dependent gravity\footnote{For the sake of completeness, we provide a detailed presentation, as in the  paper \cite{herglotz_GR2} there are several typographical errors in the derivation of the otherwise correct field equations.}.

\paragraph{Action-dependent gravity.} Consider the Lagrangian density
\begin{equation}
    \mathcal{L}=16\pi\mathcal{L}_{m}+R+\lambda_{\mu}s^{\mu},
\end{equation}
where $\lambda_{\mu}$ is an arbitrary vector field and plays the role of the dissipative coefficient.
Varying Eq.\eqref{eq32}-\eqref{eq33} yields
\begin{gather}
    \delta s^{\mu}=0 \\
    \delta(s^{\mu}\sqrt{-g})e^{-\phi}]_{,\mu}=e^{-\phi}\delta(R_{\mu\nu}g^{\mu\nu}\sqrt{-g}+16\pi\mathcal{L}_{m}\sqrt{-g}),
    \label{eq36}
\end{gather}
where we denoted $\phi:=\int \lambda_{\mu}dx^{\mu}$.

Integrating \eqref{eq36} over the region $\mathcal{V}$ 
\begin{align}
    &\int_\mathcal{V}e^{-\phi}(R_{\mu\nu}g^{\mu\nu}\delta\sqrt{-g}+R_{\mu\nu}\sqrt{-g}\delta g^{\mu\nu}+\sqrt{-g}g^{\mu\nu}\delta R_{\mu\nu})d^{n}x+\int_{\nu}e^{-\phi}\delta(16\pi\mathcal{L}_{m}\sqrt{-g})d^{n}x \notag\\
    & \int_\mathcal{V}e^{-\phi}[(R_{\mu\nu}-\frac{1}{2}g_{\mu\nu}R)\delta g^{\mu\nu}+g^{\mu\nu}\delta R_{\mu\nu}]\sqrt{-g}d^{n}x+\int_{\nu}e^{-\phi}\delta(16\pi\mathcal{L}_{m}\sqrt{-g})d^{n}x  \notag\\
    & \int_\mathcal{V}e^{-\phi}(g^{\mu\nu}\delta R_{\mu\nu}+G_{\mu\nu}\delta g^{\mu\nu})\sqrt{-g}d^{n}x+\int_{\nu}e^{-\phi}16\pi\delta(\mathcal{L}_{m}\sqrt{-g})d^{n}x=0.
    \label{eq37}
\end{align}
Separating the first term, the variation of the Ricci tensor can be calculated in the following way
\begin{align}
    \int_\mathcal{V}e^{-\phi}g^{\mu\nu}(\delta R_{\mu\nu})\sqrt{-g}d^{n}x=&\int_{\nu}d^{n}xe^{-\phi}\sqrt{-g}\delta g^{\mu\nu}[g_{\mu\nu}(\lambda^{\nu}\lambda_{\nu}-\lambda^{\nu}_{,\nu}-\Gamma^{\alpha}{}_{\gamma_\alpha}\lambda^{\gamma})-(\lambda_{\mu}\lambda_{\nu}-\lambda_{\mu,\nu}+\lambda_{\sigma}\Gamma^{\sigma}{}_{\mu\nu})] \notag \\
    =& \int_\mathcal{V}d^{n}xe^{-\phi}\sqrt{-g}\delta g^{\mu\nu}[g_{\mu\nu}(\lambda^{\nu}\lambda_{\nu}-\nabla_{\nu}\lambda^{\nu})-(\lambda_{\mu}\lambda_{\nu}-\nabla_{\nu}\lambda_{\mu})] \notag \\
    &= \int_\mathcal{V}d^{n}xe^{-\phi}\sqrt{-g}\delta g^{\mu\nu}(\Lambda_{\mu\nu}-g_{\mu\nu}\Lambda) \notag \\
    &=  \int_\mathcal{V}d^{n}xe^{-\phi}\sqrt{-g}\delta g^{\mu\nu} K_{\mu\nu},
    \label{eq38}
\end{align}
where we have defined the tensors $\Lambda_{\mu\nu}:=\frac{1}{2}(\lambda_{\mu;\nu}+\lambda_{\nu;\mu})-\lambda_{\mu}\lambda_{\nu}$, $\Lambda=\Lambda^{\mu}_{\mu}$ and $K_{\mu\nu}:=\Lambda_{\mu\nu}-g_{\mu\nu}\Lambda$.

After substituting in \eqref{eq37}, we obtain
\begin{equation}
    \int_{\mathcal{V}}e^{-\phi}\sqrt{-g}\delta g^{\mu\nu}\left(K_{\mu\nu}+G_{\mu\nu}-8\pi T_{\mu\nu}\right)d^{n}x = 0.
\end{equation}
Thus, the field equations of general relativity from the Herglotz variational principle take the form \cite{herglotz_GR2}
\begin{equation}
    K_{\mu\nu}+G_{\mu\nu}=8\pi T_{\mu\nu}.
\end{equation}
Note that in the case when $\lambda_{\mu}=0$, the equation reduces to the Einstein field equations.

It can also be seen that by taking the divergence of the field equations 
\begin{equation}
    \nabla^{\mu}(K_{\mu\nu}+G_{\mu\nu})=8\pi \nabla^{\mu}(T_{\mu\nu}),
\end{equation}
the energy-momentum tensor remains conserved if the condition $\nabla^{\mu}K_{\mu\nu}=0$ holds.

\section{Herglotz-type \texorpdfstring{$f(R,\text{Matter})$}{fRmatter} gravity}
\label{sec4}
In this section, we formulate $f(R,\text{Matter})$ theories using the Herglotz formalism in both geometric and scalar-tensor representations. We will discuss special cases, and show that our formulation recovers Herglotz-type $f(R,T)$ gravity as a special case. We also show that, in the limit as the Herglotz contribution tends to zero, $f(R,\mathcal{L}_m), f(R,T), f(R,T_{\mu\nu} T^{\mu\nu})$ theories are naturally recovered.

\subsection{Geometric representation}



The extension of the Herglotz variational principle to non-minimally coupled $f(R,\text{Matter})$ gravity follows the same general strategy as in the cases discussed previously. Hence, the Lagrangian is given by
\begin{equation}
     \mathcal{L}=16\pi\mathcal{L}_m+f(R,\Phi)+\lambda_{\mu}s^{\mu}.
\end{equation}

The metric variation gives
\begin{align}
    \frac{\delta S}{\delta g^{\mu\nu}}=&\int_{\nu}e^{-\phi}[R_{\mu\nu}f_{R}(R,\Phi)+(g_{\mu\nu}\Box-\nabla_{\mu}\nabla_{\nu})f_{R}(R,\Phi)+f_{R}(R,\Phi)K_{\mu\nu}-2g_{\mu\nu}\lambda_{\beta}\partial^{\beta}(f_{R}(R,\Phi))+\notag\\&+\lambda_{\mu}\partial_{\nu}(f_{R}(R,\Phi))+\lambda_{\nu}\partial_{\mu}(f_{R}(R,\Phi))+f_{\Phi}(R,\Phi)\Xi_{\mu\nu}-\frac{1}{2}g_{\mu\nu}f(R,\Phi)-8\pi T_{\mu\nu}]\sqrt{-g}d^{n}x=0,
\end{align}
where we denoted $K_{\mu\nu}:=\Lambda_{\mu\nu}-g_{\mu\nu}\Lambda$, $\Lambda_{\mu\nu}:=\frac{1}{2}(\lambda_{\mu;\nu}+\lambda_{\nu;\mu})-\lambda_{\mu}\lambda_{\nu}$ and $\Lambda:=\Lambda^{\mu}_{\mu}$.

Since $\delta g^{\mu\nu}$ is arbitrary, the field equations of $f(R,\text{Matter})$ within the Herglotz framework are given by
\begin{equation}
    f_{R}(R,\Phi)R_{\mu\nu}-\frac{1}{2}g_{\mu\nu}f(R,\Phi)+(g_{\mu\nu}\Box-\nabla_{\mu}\nabla_{\nu})f_{R}(R,\Phi)+H_{\mu\nu}=8\pi T_{\mu\nu}-f_{\Phi}(R,\Phi)\Xi_{\mu\nu},
    \label{eq47}
\end{equation}
where the Herglotz contribution
\begin{equation}
    H_{\mu\nu}:=f_{R}(R,\Phi)K_{\mu\nu}+\lambda_{\mu}\partial_{\nu}(f_{R}(R,\Phi))+\lambda_{\nu}\partial_{\mu}(f_{R}(R,\Phi))-2g_{\mu\nu}\lambda_{\beta}\partial^{\beta}(f_{R}(R,\Phi)).
    \label{eq48}
\end{equation}
It can be seen that in the limit of a vanishing Herglotz field, $\lambda_{\mu}=0$, Eq.\eqref{eq47} reduces to the regular $f(R,\text{Matter})$ gravity field equations.

The equations of motion for matter fields $\Psi$, are obtained from the generalized Euler-Lagrange equations 
\begin{equation}
    \frac{\partial \left(\frac{1}{16\pi}f(R,\Phi)+\mathcal{L}_{m}\right)}{\partial \Psi}
-\,\nabla_\mu\!\left(\frac{\partial \left(\frac{1}{16\pi}f(R,\Phi)+\mathcal L_{m}\right)}{\partial (\partial_\mu\Psi)}\right)+\lambda_{\mu}\frac{\partial \left(\frac{1}{16\pi}f(R,\Phi)+\mathcal L_{m}\right)}{\partial (\partial_\mu\Psi)}=0.
\label{eq52}
\end{equation}
It is worth mentioning that in the case $\Phi=\Phi(g_{\mu\nu},\Psi)$, $\mathcal{L}_{m}=\mathcal{L}_{m}(g_{\mu\nu},\Psi)$, then the equations of motion for the fields \eqref{eq52} and \eqref{eq6} coincide.

Taking the divergence of Eq.\eqref{eq47} gives
\begin{equation}
    \nabla^{\mu}(f_{R}(R,\Phi)R_{\mu\nu}-\frac{1}{2}g_{\mu\nu}f(R,\Phi)+(g_{\mu\nu}\Box-\nabla_{\mu}\nabla_{\nu})f_{R}(R,\Phi)+H_{\mu\nu})=\nabla^{\mu}(8\pi T_{\mu\nu}-f_{\Phi}(R,\Phi)\Xi_{\mu\nu}).
\end{equation}
Using the known identity $(\Box\nabla_{\nu}-\nabla_{\nu}\Box)F=R_{\mu\nu}\nabla^{\mu}F$ \cite{koivisto}, one  finds
\begin{equation}
    \nabla^{\mu}T_{\mu\nu}=\frac{1}{8\pi}\left[\nabla^{\mu}H_{\mu\nu}-\frac{1}{2}g_{\mu\nu}f_{\Phi}(R,\Phi)\nabla^{\mu}\Phi+\nabla^{\mu}(f_{\Phi}(R,\Phi)\Xi_{\mu\nu})\right].
    \label{eq50}
\end{equation}
As in the classical $f(R,\text{Matter})$ theory case, Eq.\eqref{eq50} shows that the energy-momentum tensor is generally not covariantly conserved.

This non-conservation, similarly to the classical $f(R,\text{Matter})$ case, leads to a non-geodesic motion, with an extra-force induced by the non-minimal coupling and the Herglotz contribution, given by 
\begin{equation}
\mathcal{F_H^{\mu}}:=\frac{h_{\alpha}{}^{\mu}}{8\pi(\rho+p)}\left[\left(8\pi\nabla_{\nu}p-\frac{1}{2}f_{\Phi}(R,\Phi)\nabla_{\nu}\Phi\right)g^{\alpha\nu}+\nabla_{\nu}(f_{\Phi}(R,\Phi)\Xi^{\alpha\nu})+\nabla_{\nu}H^{\mu\nu}\right],
\end{equation}
which reproduces the  extra force \eqref{eq:force} in the limit of vanishing Herglotz contribution, $H_{\mu \nu} \to 0$.

We now illustrate three special cases of interest, one reproducing the recently proposed Herglotz-type $f(R,T)$ gravity \cite{HerglotzFRT}.

\textbf{Herglotz-type $f(R,T)$ gravity.} In order to illustrate the general formalism, we choose $\Phi=T$, where $T$ denotes the trace of the energy-momentum tensor. In this case, one has
\begin{equation}
    \Xi_{\mu\nu}=\frac{\delta T}{\delta g^{\mu\nu}}=\frac{\delta (g^{\alpha\beta}T_{\alpha\beta})}{\delta g^{\mu\nu}}=g^{\alpha\beta}\frac{\delta T_{\alpha\beta}}{\delta g^{\mu\nu}}+T_{\mu\nu}=g_{\mu\nu}\mathcal{L}_{m}-T_{\mu\nu},
\end{equation}
but in literature the following form is also used
\begin{equation}\label{xifRT}
    \Xi_{\mu\nu}=\frac{\delta T}{\delta g^{\mu\nu}}=\frac{\delta (g^{\alpha\beta}T_{\alpha\beta})}{\delta g^{\mu\nu}}=\Theta_{\mu\nu}+T_{\mu\nu},
\end{equation}
where we denoted $\Theta_{\mu\nu}:=g^{\alpha\beta}\frac{\delta T_{\alpha\beta}}{\delta g^{\mu\nu}}$.

Thus, the field equations of $f(R,T)$ gravity within the Herglotz framework take the form
\begin{equation}
    f_{R}(R,T)R_{\mu\nu}-\frac{1}{2}g_{\mu\nu}f(R,T)+(g_{\mu\nu}\Box-\nabla_{\mu}\nabla_{\nu})f_{R}(R,T)+H_{\mu\nu}=8\pi T_{\mu\nu}-T_{\mu\nu}f_{T}(R,T)-\Theta_{\mu\nu}f_{T}(R,T),
    \label{eq:Herglotz f(R,T) field eq.}
\end{equation}
coinciding with the results of \cite{HerglotzFRT}.

Substituting  $\Phi=T$ and \eqref{xifRT} in Eq.\eqref{eq50}, the covariant divergence of the energy-momentum tensor is given by
\begin{align}
\nabla^{\mu}T_{\mu\nu}
=
\frac{1}{8\pi+f_{T}(R,T)}
\Biggl[
&\nabla^{\mu}H_{\mu\nu}
-\frac{1}{2}f_{T}(R,T)\nabla_{\nu}T
+(T_{\mu\nu}+\Theta_{\mu\nu})\nabla^{\mu}f_{T}(R,T)
\notag\\
&+f_{T}(R,T)\nabla_{\nu}\mathcal{L}_{m}
\Biggr],
\label{eq:EMT_Herlogtz_f(R,T)}
\end{align}
which reproduces the results of \cite{HerglotzFRT}.

\textbf{Herglotz-type $f(R,L_m)$ gravity.} In this case $\Phi=\mathcal L_m(g_{\mu\nu},\Psi,\partial_{\mu}\Psi)$, and hence the theory is given by the Lagrangian
\begin{equation}
 \mathcal{L}=f(R,\mathcal L_m(g_{\mu\nu},\Psi,\partial_{\mu}\Psi))+16\pi\mathcal L_m(g_{\mu\nu},\Psi,\partial_{\mu}\Psi)+\lambda_{\mu}s^{\mu}.
\end{equation}
Alternatively, the explicit matter term may be combined with the gravitational sector through a redefinition of the function
$\tilde f(R,\mathcal L_m):=f(R,\mathcal L_m)+16\pi\mathcal L_m$,
so that $\mathcal{L}=\tilde f(R,\mathcal L_m)+\lambda_{\mu}s^{\mu}$. Hence, we have
\begin{equation}
    \Xi_{\mu\nu}=\frac{\delta\mathcal{L}_{m}}{\delta g^{\mu\nu}}=\frac{1}{2}g_{\mu\nu}\mathcal{L}_{m}-\frac{1}{2}T_{\mu\nu}.
    \label{Xi:frlm}
\end{equation}

Substituting this in Eq.\eqref{eq47}, we get the metric field equations of $f(R,\mathcal{L}_{m})$ gravity, 
\begin{align}
    &\tilde f_{R}(R,\mathcal{L}_{m})R_{\mu\nu}+(g_{\mu\nu}\Box-\nabla_{\mu}\nabla_{\nu})\tilde f_{R}(R,\mathcal{L}_{m})-\frac{1}{2}[\tilde f(R,\mathcal{L}_{m})-\notag\\
    &-\tilde f_{\mathcal{L}_{m}}(R,\mathcal{L}_{m})\mathcal{L}_{m}]g_{\mu\nu}+H_{\mu\nu}=\frac{1}{2} \tilde f_{\mathcal{L}_{m}}(R,\mathcal{L}_{m})T_{\mu\nu}.
    \label{eq:Herglotz_f(R,Lm)fieldeq.}
\end{align}

The non-conservation law can be found by substituting the explicit form of $\Xi_{\mu \nu}$ in Eq.\eqref{eq50}, which leads to
\begin{equation}
    \nabla^{\mu}H_{\mu\nu}+ \left(\frac{1}{2}g_{\mu\nu}\mathcal{L}_{m}-\frac{1}{2}T_{\mu\nu}\right)\nabla^{\mu}(\tilde f_{\mathcal{L}_{m}}(R,\mathcal{L}_{m}))-\frac{1}{2}\tilde f_{\mathcal{L}_{m}}(R,\mathcal{L}_{m})\nabla^{\mu}T_{\mu\nu}=0,
\end{equation}
and thus, for the divergence of the energy-momentum tensor, we obtain
\begin{align}
    \nabla^{\mu}T_{\mu\nu}=\frac{2}{\tilde f_{\mathcal{L}_{m}}(R,\mathcal{L}_{m})}\left[\frac{1}{2}(g_{\mu\nu}\mathcal{L}_{m}-T_{\mu\nu})\nabla^{\mu}\tilde f_{\mathcal{L}_{m}}(R,\mathcal{L}_{m})+\nabla^{\mu}H_{\mu\nu}\right].
\end{align}

In the limit of vanishing Herglotz contribution, that is $H_{\mu\nu} \to 0$, we obtain the divergence law presented in \cite{harko_fR_Lm}.

\textbf{Herglotz-type $f(R,T_{\mu\nu} T^{\mu\nu})$ gravity.} Here $\Phi$ is given by the quadratic matter invariant, $\Phi = T_{\mu\nu}T^{\mu\nu}$, where $T_{\mu\nu}$ is the matter energy-momentum tensor. Hence, we have
\begin{align}
    \Xi_{\mu\nu}=\frac{\delta(T_{\alpha\beta}T^{\alpha\beta})}{\delta g^{\mu\nu}}=\frac{\delta(g^{\alpha\rho}g^{\beta\sigma}T_{\alpha\beta}T_{\rho\sigma})}{\delta g^{\mu\nu}}=2T^{\alpha}{}_{\mu}T_{\nu\alpha}+2T^{\alpha\beta}\frac{\delta T_{\alpha\beta}}{\delta g^{\mu\nu}}.
    \label{eq61}
\end{align}
 From Eq.\eqref{eq2}, the expression for the energy-momentum tensor is
 \begin{equation}
     T_{\alpha\beta}=g_{\alpha\beta}\mathcal{L}_{m}-2\frac{\partial \mathcal{L}_{m}}{\partial g^{\alpha\beta}},
 \end{equation}
 where we used that the Lagrangian, $\mathcal{L}_{m}$, depends only on the metric and not on its derivatives. Varying with respect to the metric leads to
 \begin{align}
     \frac{\delta T_{\alpha\beta}}{\delta g^{\mu\nu}}&=\frac{\delta g_{\alpha\beta}}{\delta g^{\mu\nu}}\mathcal{L}_{m}+g_{\alpha\beta}\frac{\partial\mathcal{L}_{m}}{\partial g^{\mu\nu}}\\
     &=-g_{\alpha\theta}g_{\beta\rho}\delta^{\theta\rho}_{\mu\nu}\mathcal{L}_{m}+\frac{1}{2}g_{\alpha\beta}g_{\mu\nu}\mathcal{L}_{m}-\frac{1}{2}g_{\alpha\beta}T_{\mu\nu},
 \end{align}
where $\delta^{\theta\rho}_{\mu\nu}:=\delta g^{\theta\rho}/\delta g^{\mu\nu}$ is the generalized Kronecker symbol.

Substituting in Eq.\eqref{eq61} yields
\begin{equation}
    \Xi_{\mu\nu}=\frac{\delta(T_{\alpha\beta}T^{\alpha\beta})}{\delta g^{\mu\nu}}=-2\mathcal{L}_{m}(T_{\mu\nu}-\frac{1}{2}g_{\mu\nu}T)-TT_{\mu\nu}+2T^{\alpha}{}_{\mu}T_{\nu\alpha}.
    \label{Xi:fTT}
\end{equation}

Plugging this back in Eq.\eqref{eq47}, we obtain the metric field equations of $f(R,T_{\mu\nu}T^{\mu\nu})$ gravity

\begin{align}
    f_{T^2}R_{\mu\nu}-\frac{1}{2}g_{\mu\nu}f+(g_{\mu\nu}\Box-\nabla_{\mu}\nabla_{\nu})f_{R}+H_{\mu\nu}=8\pi T_{\mu\nu}-f_{T^2}\left[-2\mathcal{L}_{m}(T_{\mu\nu}-\frac{1}{2}g_{\mu\nu}T)-TT_{\mu\nu}\right.+\\\left.2T^{\alpha}{}_{\mu}T_{\nu\alpha}\right],
\end{align}
where we used the notation $f_{T^2}:=\partial f / \partial(T_{\mu\nu}T^{\mu\nu})$.

Furthermore, the divergence of the energy-momentum tensor takes the form
\begin{align}
\nabla^{\mu}T_{\mu\nu}= \frac{1}{8\pi}\biggl\{
\nabla^{\mu}H_{\mu\nu}-\frac{1}{2}f_{T^2}\nabla_{\nu}(T_{\alpha\beta}T^{\alpha\beta})+\nabla^{\mu}\biggl[f_{T^2}\left(
-2\mathcal{L}_{m}(T_{\mu\nu}-\frac{1}{2}g_{\mu\nu}T)-TT_{\mu\nu}
\right.\\\left.+2T^{\alpha}{}_{\mu}T_{\nu\alpha}
\right)\biggr]\biggr\}.
\end{align}

Note that in the vanishing Herglotz field limit, $H_{\mu \nu} \to 0$, the equations of \cite{KatriciKavuk2014} are obtained.

\subsection{Scalar-tensor representation}

It is well-known, the $f(R,T)$  theory also admits a scalar-tensor representation \cite{Pinto:2023tof, Pinto:2022PhysRevD}. Recently, this was generalized to Herglotz-type $f(R,T)$ gravity in \cite{scalar_marek}. Following this approach, we formulate Herglotz-type $f(R,\text{Matter})$ gravity in the scalar-tensor representation. To this end, we introduce two auxiliary scalar fields $\alpha$ and $\beta$ which will reproduce $R$ and $\Phi$, respectively.

The corresponding Herglotz-type Lagrangian density is
\begin{equation}
\mathcal{L}
= f(\alpha,\beta) + (R-\alpha) f_{\alpha}(\alpha,\beta)
+ (\Phi-\beta) f_{\beta}(\alpha,\beta)
+ \lambda_{\mu} s^{\mu} + 16\pi \mathcal{L}_m ,
\label{eq67}
\end{equation}
where $f_{\alpha} := \partial f/\partial \alpha$ and
$f_{\beta} := \partial f/\partial \beta$.

Varying with respect to $\alpha$ and $\beta$ gives the following algebraic equations
\begin{equation}
\begin{cases}
f_{\alpha\alpha}(\alpha,\beta)(R-\alpha)
+ f_{\alpha\beta}(\alpha,\beta)(\Phi-\beta) = 0, \\[0.3em]
f_{\beta\alpha}(\alpha,\beta)(R-\alpha)
+ f_{\beta\beta}(\alpha,\beta)(\Phi-\beta) = 0.
\end{cases}
\label{eq68}
\end{equation}

Provided that the Hessian determinant is nonzero, 
\begin{equation}
\Delta := f_{\alpha\alpha} f_{\beta\beta} - f_{\alpha\beta}^{2} \neq 0 ,
\end{equation}
we find that the solution of Eq.\eqref{eq68} is
\begin{equation}
\alpha = R,
\qquad
\beta = \Phi .
\end{equation}

Introducing the scalar fields
\begin{equation}
\varphi := f_{\alpha}(\alpha,\beta),
\qquad
\psi := f_{\beta}(\alpha,\beta),
\end{equation}
together with the potential
\begin{equation}
U(\varphi,\psi) := \varphi \alpha + \psi \beta - f(\alpha,\beta),
\end{equation}
the Lagrangian density
\eqref{eq67} takes the equivalent scalar-tensor form
\begin{equation}
\mathcal{L}
= \varphi R + \psi \Phi - U(\varphi,\psi)
+ \lambda_{\mu} s^{\mu} + 16\pi \mathcal{L}_m .
\label{eq73}
\end{equation}

In this representation the nonlinear dependence on $(R,\Phi)$ is fully encoded in $U$, while $R$ and $\Phi$ enter linearly.

Varying \eqref{eq73} with respect to the metric, leads to the field equations
\begin{equation}
\varphi R_{\mu\nu}
- \frac{1}{2} g_{\mu\nu}(\varphi R + \psi \Phi - U)
+ (g_{\mu\nu}\Box - \nabla_{\mu}\nabla_{\nu})\varphi
+ H_{\mu\nu}
= 8\pi T_{\mu\nu} - \psi \Xi_{\mu\nu},
\label{eq74}
\end{equation}
where $\Xi_{\mu\nu} = \delta \Phi / \delta g^{\mu\nu}$.
The Herglotz contribution takes the form
\begin{equation}
H_{\mu\nu}
:= \varphi K_{\mu\nu}
+ \lambda_{\mu}\partial_{\nu}\varphi
+ \lambda_{\nu}\partial_{\mu}\varphi
- 2 g_{\mu\nu}\lambda^{\beta}\partial_{\beta}\varphi ,
\end{equation}
with
\[
K_{\mu\nu} = \Lambda_{\mu\nu} - g_{\mu\nu}\Lambda,
\qquad
\Lambda_{\mu\nu} = \tfrac12(\lambda_{\mu;\nu} + \lambda_{\nu;\mu})-\lambda_{\mu}\lambda_{\nu},
\qquad
\Lambda = \Lambda^{\mu}{}_{\mu}.
\]

Variation of \eqref{eq73} with respect to $\varphi$ and $\psi$ yields the scalar relations
\begin{equation}
\begin{cases}
R = \dfrac{\partial U}{\partial \varphi}, \\[0.6em]
\Phi = \dfrac{\partial U}{\partial \psi}.
\end{cases}
\label{eq76}
\end{equation}

Finally, taking the divergence of Eq.\eqref{eq74} and using \eqref{eq76} one obtains
\begin{equation}
\nabla^{\mu} T_{\mu\nu}
= \frac{1}{8\pi}
\left[
\nabla^{\mu} H_{\mu\nu}
- \frac{1}{2} g_{\mu\nu}\psi \nabla^{\mu}\Phi
+ \nabla^{\mu}(\psi \Xi_{\mu\nu})
\right].
\label{eq:Scalar_TensorEMT}
\end{equation}
The above equation is the scalar-tensor equivalent of the divergence of the energy-momentum tensor in the geometric representation \eqref{eq50}. Likewise, the energy-momentum tensor is not conserved. For sake of completeness, we also present explicitly the scalar-tensor equivalents of the three theories discussed in the geometric representation.

\textbf{Herglotz-type scalar-tensor $f(R,T)$ gravity.} Choosing $\Phi=T$, the scalar-tensor Lagrangian becomes
\begin{equation}
\mathcal{L}=\varphi R+\psi T-U(\varphi,\psi)+\lambda_\mu s^\mu+16\pi \mathcal{L}_m .
\end{equation}
Using Eq.\eqref{xifRT}, the field equations are given by
\begin{equation}
\phi R_{\mu\nu}
-\frac12 g_{\mu\nu}(\phi R+\psi T-U)
+(g_{\mu\nu}\Box-\nabla_\mu\nabla_\nu)\phi
+H_{\mu\nu}
=
8\pi T_{\mu\nu}
-\psi(T_{\mu\nu}+\Theta_{\mu\nu}),
\end{equation}
The covariant divergence of the energy-momentum tensor can be obtained by substituting in Eq.\eqref{eq:Scalar_TensorEMT}
\begin{equation}
\nabla^\mu T_{\mu\nu}=\frac{1}{8\pi}\left[\nabla^\mu H_{\mu\nu}
-\frac{1}{2}\psi\nabla_\nu T
+\nabla^\mu\bigl(\psi(T_{\mu\nu}+\Theta_{\mu\nu})\bigr)
\right],
\end{equation}
or equivalently
\begin{equation}
    \nabla^\mu T_{\mu\nu}=\frac{1}{8\pi+\psi}\left[\nabla^\mu H_{\mu\nu}+\frac{1}{2}\psi\nabla_\nu(2\mathcal{L}_{m}-T)
+(T_{\mu\nu}+\Theta_{\mu\nu})\nabla^\mu\psi
\right],
\end{equation}
which gives back the results of \cite{scalar_marek}.

\textbf{Herglotz-type scalar-tensor $f(R,\mathcal{L}_m)$ gravity.} For $\Phi=\mathcal{L}_m$, the scalar-tensor Lagrangian reads
\begin{equation}
L=\varphi R+\psi \mathcal{L}_m-U(\phi,\psi)+\lambda_\mu s^\mu,
\end{equation}
where we neglected the $16\pi \mathcal{L}_m$ term.

Using the relation \eqref{Xi:frlm}, after subsituting in Eq.\eqref{eq74} we obtain the metric field equations
\begin{equation}
\varphi R_{\mu\nu}
-\frac12 g_{\mu\nu}(\varphi R-U)
+(g_{\mu\nu}\Box-\nabla_\mu\nabla_\nu)\varphi+H_{\mu\nu}
=\frac{1}{2}\psi T_{\mu\nu}.
\end{equation}
Taking into account that we neglected the $16\pi\mathcal{L}_{m}$ and using Eq.\eqref{eq:Scalar_TensorEMT}, the non-conservation law is given by
\begin{equation}
\nabla^\mu T_{\mu\nu}
=
\frac{2}{\psi}
\left[
\nabla^\mu H_{\mu\nu}+\frac{1}{2}\left(g_{\mu\nu}\mathcal{L}_{m}-T_{\mu\nu}\right)\nabla^{\mu}\psi
\right].
\end{equation}

\textbf{Herglotz-type scalar-tensor $f(R,T_{\mu\nu} T^{\mu \nu})$ gravity.} For the quadratic matter invariant $\Phi=T_{\alpha\beta}T^{\alpha\beta}$,
the scalar-tensor Lagrangian is
\begin{equation}
\mathcal{L}=\varphi R+\psi T_{\alpha\beta}T^{\alpha\beta}
-U(\phi,\psi)+\lambda_\mu s^\mu+16\pi \mathcal{L}_m .
\end{equation}
Using the previous result \eqref{Xi:fTT}, the field equations become
\begin{align}
\varphi R_{\mu\nu}&-\frac12 g_{\mu\nu}
\left(\varphi R+\psi T_{\alpha\beta}T^{\alpha\beta}-U
\right)+(g_{\mu\nu}\Box-\nabla_\mu\nabla_\nu)\varphi
+H_{\mu\nu}\nonumber\\&=8\pi T_{\mu\nu}-\psi\Bigg[
-2L_m\left(T_{\mu\nu}-\frac12 g_{\mu\nu}T\right)-TT_{\mu\nu}
+2T^\alpha{}_\mu T_{\nu\alpha}
\Bigg].
\end{align}

Finally, the covariant divergence of the energy-momentum tensor can be obtained by substituting in Eq.\eqref{eq:Scalar_TensorEMT}
\begin{align}
\nabla^\mu T_{\mu\nu}
=
\frac{1}{8\pi}
\Bigg\{
&\nabla^\mu H_{\mu\nu}
-\frac12 \psi \nabla_\nu
\left(T_{\alpha\beta}T^{\alpha\beta}\right)
\nonumber\\
&+\nabla^\mu\Bigg[
\psi\Bigg(
-2\mathcal{L}_m\left(T_{\mu\nu}-\frac12 g_{\mu\nu}T\right)
-TT_{\mu\nu}
+2T^\alpha{}_\mu T_{\nu\alpha}
\Bigg)
\Bigg]
\Bigg\}.
\end{align}


\subsection{Preserving the energy-momentum tensor in Herglotz-type \texorpdfstring{$f(R,\text{Matter})$}{fr} theories}
\label{sec5}

In the Herglotz formalism, the dissipation one-form, which is closed, does not possess dynamics a priori. Therefore, to fully fix the theory, one should either choose a closed one-form with components $\lambda_{\mu}=\partial_{\mu} f$ for some scalar field $f$, or prescribe dynamics, which determine $\lambda_{\mu}$. In this section, we show that for any $f(R,\text{Matter})$ theory there is a natural way to fix the Herglotz contribution, such that the matter energy-momentum tensor becomes covariantly conserved.

From the divergence equation 
\begin{equation}
    \nabla^{\mu}T_{\mu\nu}=\frac{1}{8\pi}\left[\nabla^{\mu}H_{\mu\nu}-\frac{1}{2}g_{\mu\nu}f_{\Phi}(R,\Phi)\nabla^{\mu}\Phi+\nabla^{\mu}(f_{\Phi}(R,\Phi)\Xi_{\mu\nu})\right],
\end{equation}
we see that if the Herglotz contribution $H_{\mu\nu}$ obeys
\begin{equation}
    \nabla^{\mu}H_{\mu\nu}=\frac{1}{2}g_{\mu\nu}f_{\Phi}(R,\Phi)\nabla^{\mu}\Phi-\nabla^{\mu}(f_{\Phi}(R,\Phi)\Xi_{\mu\nu}),
    \label{eq51}
\end{equation}
then the energy-momentum tensor becomes conserved. 

This can be equivalently formulated in the scalar-tensor representation, considering Eq.\eqref{eq:Scalar_TensorEMT}, the condition is given by
\begin{equation}
\nabla^{\mu} H_{\mu\nu}
= \frac{1}{2} g_{\mu\nu}\psi \nabla^{\mu}\Phi
- \nabla^{\mu}(\psi \Xi_{\mu\nu}),
\end{equation}
for non-trivial cases with $\psi = f_{\beta} \neq 0$.

A consequence of the induced Herglotz condition, Eq.\eqref{eq51}. is that in this case the extra-force $\mathcal{F}^{\mu}_{\mathcal{H}}$ acting on a test fluid, simplifies to 
\begin{equation}
    \mathcal{F_H^{\mu}}-\frac{h^{\mu\nu}\nabla_{\nu}p}{(\rho+p)}=0.
\end{equation}
After substitution, we find that the equation of motion for perfect fluids in Herglotz-type $f(R,\text{Matter})$ theories match the ones from classical general relativity, that is 
\begin{equation}
    \frac{d^2x^{\mu}}{ds^2}+\Gamma^{\mu}{}_{\nu\lambda}u^{\nu}u^{\lambda}=\frac{h^{\mu\nu}\nabla_{\nu}p}{(\rho+p)}.
\end{equation}

The explicit conditions on the Herglotz field that guarantee covariant energy-momentum conservation for the previously discussed theories are summarized in Table \ref{table1}.

\begin{table*}[t]
\centering
\footnotesize
\setlength{\tabcolsep}{6pt}
\renewcommand{\arraystretch}{2.0}
\begin{tabularx}{\textwidth}{
|>{\centering\arraybackslash}m{0.16\textwidth}
|>{\raggedright\arraybackslash}X
|>{\raggedright\arraybackslash}X|
}
\hline
\textbf{Theory} &
\textbf{Geometric representation} &
\textbf{Scalar-tensor representation}
\\
\hline

$f(R,\Phi)$
&
$\nabla^\mu H_{\mu\nu}
= \dfrac{1}{2}f_\Phi \nabla_\nu\Phi
- \nabla^\mu(f_\Phi \Xi_{\mu\nu})$
&
$\nabla^\mu H_{\mu\nu}
= \dfrac{1}{2}\psi \nabla_\nu\Phi
- \nabla^\mu(\psi \Xi_{\mu\nu})$
\\[1.3em]
\hline

$f(R,T)$
&
$\begin{aligned}
\nabla^\mu H_{\mu\nu}
={}& -(\Theta_{\mu\nu}+T_{\mu\nu})\nabla^\mu f_T \\
&+ \dfrac{1}{2}f_T \nabla_\nu(T-2\mathcal{L}_m)
\end{aligned}$
&
$\begin{aligned}
\nabla^\mu H_{\mu\nu}
={}& \dfrac{1}{2}\psi \nabla_\nu(T-2\mathcal{L}_m) \\
&- (T_{\mu\nu}+\Theta_{\mu\nu})\nabla^  \mu\psi
\end{aligned}$
\\[1.3em]
\hline

$f(R,\mathcal{L}_m)$
&
$\nabla^\mu H_{\mu\nu}
= \dfrac{1}{2}(T_{\mu\nu}-g_{\mu\nu}\mathcal{L}_m)
\nabla^\mu \tilde f_{L_m}$
&
$\nabla^\mu H_{\mu\nu}
= \dfrac{1}{2}(T_{\mu\nu}-g_{\mu\nu}\mathcal{L}_m)
\nabla^\mu \psi$
\\[1.3em]
\hline

$f(R,T_{\alpha\beta}T^{\alpha\beta})$
&
$\begin{aligned}
\nabla^\mu H_{\mu\nu}
&= \dfrac{1}{2}f_{T^2}
\nabla_\nu(T_{\alpha\beta}T^{\alpha\beta}) \\
&- \nabla^\mu\!\left[
f_{T^2}\!\left(
-2\mathcal{L}_m\!\left(T_{\mu\nu}
-\dfrac{1}{2}g_{\mu\nu}T\right)\right.\right. \\
&\left.\left.
- TT_{\mu\nu}
+ 2T^\alpha{}_\mu T_{\nu\alpha}
\right)\right]
\end{aligned}$
&
$\begin{aligned}
\nabla^\mu H_{\mu\nu}
&= \dfrac{1}{2}\psi
\nabla_\nu(T_{\alpha\beta}T^{\alpha\beta}) \\
&- \nabla^\mu\!\left[
\psi\!\left(
-2\mathcal{L}_m\!\left(T_{\mu\nu}
-\dfrac{1}{2}g_{\mu\nu}T\right)\right.\right. \\
&\left.\left.
- TT_{\mu\nu}
+ 2T^\alpha{}_\mu T_{\nu\alpha}
\right)\right]
\end{aligned}$
\\
\hline

\end{tabularx}
\caption{Summary of Herglotz-type $f(R,\mathrm{Matter})$ theories and the corresponding conservation conditions in the geometric and scalar-tensor representations.}
\label{table1}
\end{table*}

\section{Discussion and final remarks}
\label{sec6}

In this work, we developed a unified description of gravitational theories in which the gravitational Lagrangian depends on the Ricci scalar and a general matter scalar invariant, $f(R,\Phi)$, where $\Phi$ may represent, for example, $T,\mathcal{L}_{m}, T_{\mu\nu} T^{\mu\nu}$.  This formulation describes several common theories with non-minimal coupling in a compact way. The main consequence of the non-minimal coupling  is the non-conservation of the energy-momentum tensor. In the standard formulation, this is expressed as
\begin{equation}
\nabla_\mu T^{\mu\nu} = J^\nu ,
\end{equation}
where 
\begin{equation}
J^\nu :=
\frac{1}{8\pi}
\left[
-\frac{1}{2} f_\Phi(R,\Phi)\nabla^\nu \Phi
+ \nabla_\mu \left( f_\Phi(R,\Phi)\Xi^{\mu\nu} \right)
\right],
\end{equation}
The current $J_\nu$ measures the exchange between the matter sector and the additional geometric structure induced by the non-minimal coupling. If $J_\nu\neq 0$, the motion of matter is modified with respect to general relativity. For a perfect fluid this gives rise to an extra force, and in the pressureless limit test particles do not generally follow geodesics.

We then extended the theory by using the Herglotz variational principle. In this formulation the action becomes dynamical, and the theory contains a closed one-form $\lambda_\mu$, which enters the field equations through the Herglotz contribution $H_{\mu\nu}$. This contribution modifies the balance equation, which now takes the form
\begin{equation}
\nabla_\mu T^{\mu\nu} = J^{(H)}_{\nu},
\end{equation}
with
\begin{equation}
J^{(H)}_{\nu} :=
\frac{1}{8\pi}
\left[
\nabla_\mu H^{\mu\nu}
-\frac{1}{2} f_\Phi(R,\Phi)\nabla^\nu \Phi
+ \nabla_\mu \left( f_\Phi(R,\Phi)\Xi^{\mu\nu} \right)
\right].
\end{equation}
The main result of the paper is that the Herglotz one-form can be constrained so that the non-conservation induced by the non-minimal coupling is exactly compensated. More precisely, if
\begin{equation}
\nabla^\mu H_{\mu\nu}
=
\frac{1}{2} f_\Phi(R,\Phi)\nabla_\nu\Phi
-
\nabla^\mu\left(f_\Phi(R,\Phi)\Xi_{\mu\nu}\right),
\label{eq:discussion_condition}
\end{equation}
then we get
\begin{equation}
\nabla^\mu T_{\mu\nu}=J^{(H)}_\nu = 0.\label{eq103}
\end{equation}
Therefore, even in the presence of a nontrivial matter--geometry coupling, the matter energy-momentum tensor can be preserved by a suitable choice of the Herglotz contribution. In this sense, the Herglotz sector acts as a geometric compensating mechanism for the exchange current generated by the non-minimal coupling.

This result also has a direct physical consequence. When condition \eqref{eq:discussion_condition} is satisfied, the extra force generated by the non-minimal coupling is cancelled by the Herglotz contribution. The equation of motion of a perfect fluid then reduces to the same form as in general relativity,
\begin{equation}
\frac{d^2x^\mu}{ds^2}
+
\Gamma^\mu{}_{\nu\lambda}u^\nu u^\lambda
=
\frac{h^{\mu\nu}\nabla_\nu p}{\rho+p}.
\end{equation}
Consequently, for pressureless matter, $p=0$, the motion becomes geodesic. Hence, the Herglotz extension provides a way to keep non-minimal matter--geometry couplings while restoring the usual conservation law of the matter sector.

The condition \eqref{eq:discussion_condition} should not be interpreted as an automatic identity. Rather, it is a dynamical restriction on the Herglotz one-form $\lambda_\mu$, since $H_{\mu\nu}$ is built from $\lambda_\mu$, its derivatives, and $f_R$. Therefore, the preservation of the energy-momentum tensor fixes part of the freedom present in the Herglotz sector. The existence and uniqueness of solutions for $\lambda_\mu$ depend on the geometry, the chosen function $f(R,\Phi)$, the matter source, and the imposed boundary conditions. This issue deserves further investigation.

We also showed that several known theories are recovered as particular cases. For $\Phi=T$, the formalism reproduces Herglotz-type $f(R,T)$ gravity. For $\Phi=\mathcal{L}_m$, one obtains a Herglotz extension of $f(R,\mathcal{L}_m)$ gravity. For $\Phi=T_{\mu\nu}T^{\mu\nu}$, the same procedure gives the corresponding Herglotz-type extension of energy-momentum-squared gravity. In each case, the standard non-conservative theory is recovered in the limit of vanishing Herglotz contribution, $H_{\mu\nu}\to 0$.

Several future directions of research remain to be investigated. First, it would be worth exploring whether the Herglotz modifications would lead to viable cosmological models, and late-time acceleration, as was the case for $f(R,T)$ gravity \cite{HerglotzFRT,scalar_marek}. In particular, several simple $f(R,T)$ models, formulated with the usual variational principle, yield a constant deceleration parameter, and hence are unrealistic. However, in the Herglotz formalism, this problem is cured. After studying the background evolution, it would also be interesting to formulate a general setting for perturbations of a wide class of theories within the Herglotz framework, as was done for simple action-dependent gravity in \cite{herglotz_GR2}. Compact objects could also give a way to test the theory, and it would be interesting to see whether similar solutions to those obtained in \cite{PhysRevD.110.124056, PhysRevD.99.124031} exist in $f(R,\text{Matter})$ theories.

In conclusion, Herglotz-type $f(R,\text{Matter})$ gravity provides a general framework in which non-minimal matter--geometry couplings and covariant conservation of the matter energy-momentum tensor can coexist. The key mechanism is the Herglotz contribution, whose divergence can be chosen to cancel the exchange current produced by the non-minimal coupling. This opens a new class of conservative modified gravity models, including conservative versions of $f(R,T)$, $f(R,\mathcal{L}_m)$, and $f(R,T_{\mu\nu}T^{\mu\nu})$ gravity.

\section*{ACKNOWLEDGMENTS}
The author thanks Lehel Csillag for valuable discussions and helpful suggestions that improved the quality of the manuscript.

\bibliographystyle{unsrt}
\bibliography{references}

@article{koivisto,
    author = "Koivisto, Tomi",
    title = "{Covariant conservation of energy momentum in modified gravities}",
    eprint = "gr-qc/0505128",
    archivePrefix = "arXiv",
    doi = "10.1088/0264-9381/23/12/N01",
    journal = "Class. Quant. Grav.",
    volume = "23",
    pages = "4289--4296",
    year = "2006"
}

@article{HerglotzFRT,
  author        = {Wazny, Marek and Csillag, Lehel and Pinto, Miguel A. S. and Harko, Tiberiu},
  title         = {Herglotz-type $f(R,T)$ gravity},
  year          = {2025},
  eprint        = {2511.16304},
  archivePrefix = {arXiv},
  primaryClass  = {gr-qc}
}

@article{harko_fR_T,
  author = {Harko, Tiberiu and Lobo, Francisco S. N. and Nojiri, Shin'ichi and Odintsov, Sergei D.},
  title = {f(R,T) gravity},
  journal = {Physical Review D},
  volume = {84},
  number = {2},
  pages = {024020},
  year = {2011},
  publisher = {APS},
  doi = {10.1103/PhysRevD.84.024020},
  archivePrefix = {arXiv},
  eprint = {1104.2669},
}

@article{scalar_marek,
	author={Wazny, Marek},
	title={Conservative cosmology in scalar-tensor Herglotz f(R,T) gravity},
	journal={Classical and Quantum Gravity},
	url={http://iopscience.iop.org/article/10.1088/1361-6382/ae49db},
	year={2026},
}

@Article{universe11120386,
AUTHOR = {Olmo, Gonzalo J. and Pinto, Miguel A. S.},
TITLE = {On Energy-Momentum Conservation in Non-Minimal Geometry-Matter Coupling Theories},
JOURNAL = {Universe},
VOLUME = {11},
YEAR = {2025},
NUMBER = {12},
ARTICLE-NUMBER = {386},
URL = {https://www.mdpi.com/2218-1997/11/12/386},
ISSN = {2218-1997},
DOI = {10.3390/universe11120386}
}

@article{Akarsu2020RastallLCDM,
  author  = {Akarsu, {\"O}zg{\"u}r and Kat{\i}rc{\i}, Nihan and Kumar, Suresh and Nunes, Rafael C. and {\"O}zt{\"u}rk, Burcu and Sharma, Shivani},
  title   = {Rastall gravity extension of the standard {$\Lambda$}CDM model: theoretical features and observational constraints},
  journal = {The European Physical Journal C},
  year    = {2020},
  volume  = {80},
  number  = {11},
  pages   = {1050},
  doi     = {10.1140/epjc/s10052-020-08586-4}
}

@article{DUBEY2025116938,
title = {Investigating the H0-rd tension in f(R,T) gravity using cosmological observations},
journal = {Nuclear Physics B},
volume = {1017},
pages = {116938},
year = {2025},
issn = {0550-3213},
doi = {https://doi.org/10.1016/j.nuclphysb.2025.116938},
url = {https://www.sciencedirect.com/science/article/pii/S0550321325001476},
author = {Shraddha Dubey and Aroonkumar Beesham and Değer Sofuoğlu and Bhupendra Kumar Shukla and Sudha Agrawal}
}

@article{Sahlu2024CosmologyEPJC,
  author  = {Sahlu, Shambel and Alfedeel, Alnadhief H. A. and Abebe, Amare},
  title   = {The cosmology of $f(R, L_m)$ gravity: constraining the background and perturbed dynamics},
  journal = {European Physical Journal C},
  volume  = {84},
  pages   = {982},
  year    = {2024},
  doi     = {10.1140/epjc/s10052-024-13307-2}
}

@book{landau1975,
  author = {Landau, L. D. and Lifshitz, E. M.},
  title = {The Classical Theory of Fields},
  publisher = {Pergamon Press},
  year = {1975},
  edition = {4th},
  series = {Course of Theoretical Physics},
  volume = {2}
}

@article{BarroseSa:2025uxe,
    author = "Barros e S{\'a}, Nuno and Pinto, Miguel A. S. and Trindade, Tom{\'a}s",
    title = "{Clarifying the relation between covariantly conserved currents and Noether's second theorem}",
    eprint = "2506.14454",
    archivePrefix = "arXiv",
    primaryClass = "gr-qc",
    month = "6",
    year = "2025"
}

@article{herglotz_mathematical,
  author = {Lazo, Matheus J. and Paiva, L. C. T. and Amaral, J. T. S. and Frederico, Gastao S. F.},
  title = {A gauge-invariant dissipative variational principle for classical mechanics},
  journal = {Journal of Mathematical Physics},
  volume = {59},
  number = {3},
  pages = {032902},
  year = {2018},
  doi = {10.1063/1.5009682},
  archivePrefix = {arXiv},
  eprint = {1708.05770},
}

@article{herglotz_examples,
author = {Lazo, Matheus and Paiva, Juilson and Amaral, João and Frederico, Gastao},
year = {2018},
month = {03},
pages = {032902},
title = {An Action Principle for Action-dependent Lagrangians: toward an Action Principle to non-conservative systems},
volume = {59},
journal = {Journal of Mathematical Physics},
doi = {10.1063/1.5019936}
}

@article{herglotz_GR1,
    author = "Lazo, Matheus J. and Paiva, Juilson and Amaral, Jo\~ao T. S. and Frederico, Gast\~ao S. F.",
    title = "{Action principle for action-dependent Lagrangians toward nonconservative gravity: Accelerating universe without dark energy}",
    eprint = "1705.04604",
    archivePrefix = "arXiv",
    primaryClass = "gr-qc",
    doi = "10.1103/PhysRevD.95.101501",
    journal = "Phys. Rev. D",
    volume = "95",
    number = "10",
    pages = "101501",
    year = "2017"
}

@article{herglotz_GR2,
    author = "Paiva, Juilson A. P. and Lazo, Matheus J. and Zanchin, Vilson T.",
    title = "{Generalized nonconservative gravitational field equations from Herglotz action principle}",
    eprint = "2108.02902",
    archivePrefix = "arXiv",
    primaryClass = "gr-qc",
    doi = "10.1103/PhysRevD.105.124023",
    journal = "Phys. Rev. D",
    volume = "105",
    number = "12",
    pages = "124023",
    year = "2022"
}

@article{harko_fR_Lm,
  author        = {Harko, Tiberiu and Lobo, Francisco S. N.},
  title         = {{$f(R,L_m)$ gravity}},
  journal       = {The European Physical Journal C},
  volume        = {70},
  pages         = {373--379},
  year          = {2010},
  doi           = {10.1140/epjc/s10052-010-1467-3},
  eprint        = {1008.4193},
  archivePrefix = {arXiv},
  primaryClass  = {gr-qc}
}

@article{Sotiriou:2008rp,
    author = "Sotiriou, Thomas P. and Faraoni, Valerio",
    title = "{f(R) Theories Of Gravity}",
    eprint = "0805.1726",
    archivePrefix = "arXiv",
    primaryClass = "gr-qc",
    doi = "10.1103/RevModPhys.82.451",
    journal = "Rev. Mod. Phys.",
    volume = "82",
    pages = "451--497",
    year = "2010"
}

@article{Pinto:2023tof,
    author = "Pinto, Miguel A. S. and Harko, Tiberiu and Lobo, Francisco S. N.",
    title = "{Challenging \(\Lambda \)CDM with Scalar-tensor \(f(R,T)\) Gravity and Thermodynamics of Irreversible Matter Creation}",
    doi = "10.5506/APhysPolBSupp.16.6-A28",
    journal = "Acta Phys. Polon. Supp.",
    volume = "16",
    number = "6",
    pages = "28",
    year = "2023"
}

@article{Pinto:2022PhysRevD,
  author  = {Pinto, Miguel A. S. and Harko, Tiberiu and Lobo, Francisco S. N.},
  title   = {Gravitationally induced particle production in scalar-tensor $f(R,T)$ gravity},
  journal = {Physical Review D},
  volume  = {106},
  pages   = {044043},
  year    = {2022},
  doi     = {10.1103/PhysRevD.106.044043}
}

@article{KatriciKavuk2014,
  author  = {Katırcı, Nihan and Kavuk, Mehmet},
  title   = {$f(R, T_{\mu\nu}T^{\mu\nu})$ gravity and Cardassian-like expansion as one of its consequences},
  journal = {The European Physical Journal Plus},
  volume  = {129},
  pages   = {163},
  year    = {2014},
  doi     = {10.1140/epjp/i2014-14163-6}
}

@article{PhysRevD.110.124056,
  title = {Characterization of wormhole spacetimes supported by a covariant action-dependent Lagrangian theory},
  author = {Ayuso, Ismael and Lazkoz, Ruth},
  journal = {Phys. Rev. D},
  volume = {110},
  issue = {12},
  pages = {124056},
  numpages = {15},
  year = {2024},
  month = {Dec},
  publisher = {American Physical Society},
  doi = {10.1103/PhysRevD.110.124056},
  url = {https://link.aps.org/doi/10.1103/PhysRevD.110.124056}
}

@article{PhysRevD.99.124031,
  title = {Existence of static spherically-symmetric objects in action-dependent Lagrangian theories},
  author = {Fabris, Julio C. and Velten, Hermano and Wojnar, Aneta},
  journal = {Phys. Rev. D},
  volume = {99},
  issue = {12},
  pages = {124031},
  numpages = {6},
  year = {2019},
  month = {Jun},
  publisher = {American Physical Society},
  doi = {10.1103/PhysRevD.99.124031},
  url = {https://link.aps.org/doi/10.1103/PhysRevD.99.124031}
}

\end{document}